%% file: ms.tex
\documentclass[sigconf]{acmart}
\usepackage{times}
\usepackage{helvet}
\usepackage{courier}
\usepackage{graphicx}
\usepackage{algorithm}
\usepackage{algorithmic}
\usepackage{tabu}
\usepackage{array}
\usepackage{wrapfig}
\usepackage{dblfloatfix}
\usepackage[section]{placeins}
\def\BibTeX{{\rm B\kern-.05em{\sc i\kern-.025em b}\kern-.08emT\kern-.1667em\lower.7ex\hbox{E}\kern-.125emX}}

\begin{document}

\title{Debiasing Community Detection: The Importance of Lowly-Connected Nodes}

\author{Ninareh Mehrabi}
\affiliation{%
  \institution{USC Information Sciences Institute }
}
\email{ninarehm@usc.edu}

\author{Fred Morstatter}
\affiliation{%
  \institution{USC Information Sciences Institute}
}
\email{fredmors@isi.edu}

\author{Nanyun Peng}
\affiliation{%
  \institution{USC Information Sciences Institute}
}
\email{npeng@isi.edu}

\author{Aram Galstyan}
\affiliation{%
  \institution{USC Information Sciences Institute}
}
\email{galstyan@isi.edu}
 
\renewcommand{\shortauthors}{Mehrabi, et al.}
\begin{abstract}
Community detection is an important task in social network analysis, allowing us to identify and understand the communities within the social structures. However, many community detection approaches either fail to assign low degree (or lowly-connected) users to communities, or assign them to trivially small communities that prevent them from being included in analysis. In this work, we investigate how excluding these users can bias analysis results. We then introduce an approach that is more inclusive for lowly-connected users by incorporating them into larger groups. Experiments show that our approach outperforms the existing state-of-the-art in terms of F1 and Jaccard similarity scores while reducing the bias towards low-degree users. 
\end{abstract}
\maketitle
\input{sections/intro.tex}

\input{sections/relwork.tex}

\input{sections/datasets.tex}

\input{sections/approach.tex}

\input{sections/cd-expt.tex}

\input{sections/synth-expt.tex}

\input{sections/conclusions.tex}

\bibliographystyle{ACM-Reference-Format.bst}
%\bibliography{references}

\end{document}

%% file: sections/intro.tex
\vspace{-1pt}
\section{Introduction}
\begin{figure}[H]
  \centering
  \includegraphics[width=0.28\textwidth]{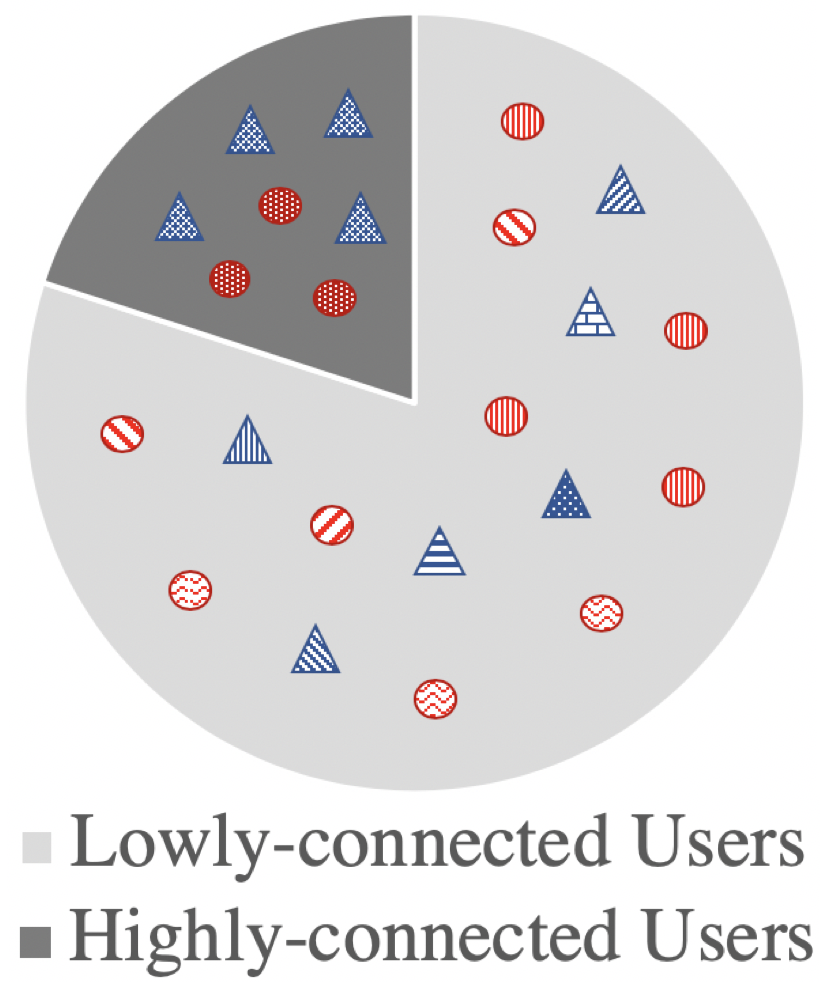}
  \caption{Demonstration of versatility of relevant low-degree people in social networks who get ignored by biased community detection methods. Shapes (circle, triangle) represent opinion, and texture represents variations of the opinion. We provide real-world examples of how this manifests in subsequent sections.}
  \label{pi}
\end{figure}

Community detection is a fundamental task in social network analysis~\cite{tang2010community}, which identifies sub-groups within social networks. These groups can represent a variety of things including karate club membership, political leanings, and deeply-held beliefs. Traditionally, these groups are identified by searching for densely-connected groups of nodes in the graph~\cite{malliaros2013clustering}. More recently, attributed approaches go beyond merely the links to cluster nodes based upon their attributes and their network connections~\cite{ruan2013efficient}. Existing community detection approaches suffer from a major flaw: the inability to assign lowly-connected users into communities. Despite the fact that lowly-connected users are not well-blended into the social network, the information they provide can be crucial for better understanding the motivations and beliefs of the community, especially considering there could be a long-tail of lowly-connected users. Failure to incorporate the lowly-connected users may result in biased results. For instance, studying groups within a social network, such as a Twitter retweet network, can be biased towards users whose tweets get a large number of retweets. This will lead to a biased analysis of the data in which not all users' voices are heard. Instead, those high-degree users with non-genuine retweets will end up having more voice in the study. Consider the toy example shown in Figure~\ref{pi}. Each node represents a user and the shapes represent difference of opinion in the network; the shade and different texture of each shape represent their variability of expressing opinions. Traditional community detection methods that solely focus on the highly-connected nodes would only capture the opinions of users within the dark-grey area; however, many diverse opinions are lost due to the fact that the users in the light grey area are excluded because of their low degree in the network.

The goal of this paper is twofold. First, we demonstrate empirically the existence of biases in existing community detection approaches using real-world datasets. Through the analysis, we show that current state-of-the-art community detection methods suffer from ignoring low-degree users that have few links either by failing to incorporate them, or by putting them into small groups which are then ignored in the study. These low-degree users have value to be included in the communities, as they offer a more diverse, nuanced representation of the communities. Second, to overcome this issue, we introduce a new community detection method, Communities with Lowly-connected Attributed Nodes (CLAN), that would mitigate the existence of this bias towards low-degree nodes.
%We prove our statement by conducting an experiment in which the state of the art methods are compared to a ground truth that was collected using human annotators and show the superiority of our method in terms of Jaccard similarity and F1 scores. To prove our point further, we use another dataset with a ground truth collected in a more automated fashion and demonstrate CLAN's ability in improving the results. 
Our contributions are as follows:
\begin{table*}
\begin{tabular}{ |p{1.5cm}||p{1.8cm}|p{5.3cm}||p{2cm}|p{4.7cm}|  }
 \hline
 \multicolumn{1}{|c||}{}&\multicolumn{2}{c||}{Gamergate Dataset}&\multicolumn{2}{c|}{U.S. Presidential Election Dataset}\\
 \hline
Method & Discarded Unique Hashtags &Examples& Discarded Unique Hashtags &Examples\\
 \hline
 CESNA   &597 (38.6\%)& \#gamerignorance,\#ethicsinjournalism, \#feministgate,\#sexismiswickedcool   & 1518 (66.9\%)&\#refugeeswelcome, \#Guns,\#OscarsStillSoWhite, \#TrumpDumped\\
 \hline
 Modularity&   625 (40.4\%) &\#misandricfeminists, \#violenceagainstwomen,\#gamersagainstgamergate, \#stopgamejournalism2014,  &302 (13.3\%)&\#Christians4Hillary,\#Killary, \#FakeHateCrimes, \#TrumpRiots \\
 \hline
\end{tabular}
\caption{Number of missed hashtags which can be representative of missing information along with some examples in biased methods.}
\label{hashtags}
\end{table*}
\begin{enumerate}
\item We introduce CLAN, a novel community detection approach that is able to categorize low-degree users into their relevant communities.
\item Through experimentation, we show that CLAN is able to outperform the existing state-of-the-art community detection methods in terms of predictive accuracy while still being able to classify more users. We conduct further experiments to test the resiliency of CLAN to different data regimes.
\item We demonstrate the existence of bias in community detection approaches that is introduced from ignoring low-degree users. We show that CLAN is able to overcome this challenge by classifying low-degree users. 

\end{enumerate}
\section{Biases in Community Detection Methods}
Analysis on community detection tends to focus on the largest communities. Methods that tend to exclude low-degree nodes are at greater risk of losing information in their detected significant communities. 
We demonstrate the existence of this bias by showing what is omitted by existing community detection approaches. We use two state-of-the-art approaches, CESNA~\cite{6729613}, and the Louvain method~\cite{blondel2008fast}. Louvain uses only the network while assigning communities, while CESNA uses both the network and user attributes. We apply these methods to two separate datasets, one based on Gamergate and one based on the 2016 U.S. Presidential Election. Both of these datasets contain two major contingents, and ground truth information (both datasets, and the methodology for obtaining ground truth, will be introduced in detail later). Table~\ref{hashtags} shows the information that is omitted by excluding the lowly-connected users. These statistics are taken by comparing the ground truth labels with the labels assigned by the community detection algorithm.
For example, in one of the methods, CESNA, %FM: which one?
the ``\#refugeeswelcome'' hashtag and the user who tweeted this hashtag remained unlabeled by the user being put in a small insignificant community from the election dataset, discussed in the later sections. By not including this user tweeting this hashtag, we lose this piece of information when we analyze the communities. In other words, degree is not correlated with relevancy to the topic and therefore merely degree and connections in social networks should not be indicative of community membership. Other examples of such are demonstrated in Table~\ref{hashtags}, such as the ``\#Christians4Hillary,'' ``\#gamersagainstgamergate,'' and many other hashtags of such type in two of the datasets that we will be using in this paper. The goal of this work is to introduce a method that would mitigate this bias by the inclusion of low-degree users into their correct belonging significant communities.

%% file: sections/relwork.tex
\vspace{-1pt}
\section{Related Work}
The related work will be discussed from two different perspectives. First, we will discuss network sampling bias, and the consequences thereof. Next, we will introduce state-of-the-art community detection approaches with a special focus on those that we use as baselines in this work.

There is a body of work surrounding different types of biases in social networks, such as the retweet network~\cite{Diaz:2018:AAB:3173574.3173986,Liao:2016:SUB:2858036.2858422}. However, these studies do not specify the root of the existing biases in social networks and ways we can mitigate them through community detection. In this paper, we show that the root of existing biases in social networks come from not only the network structure but also community detection methods that exemplify bias by putting lowly-connected nodes into non-significant communities which leads to their ignorance in the downstream studies of these communities. Therefore, we analyze some state of the art community detection methods and investigate their flaws and weaknesses. 

Current community detection methods, such as relying on the modularity value in network structures~\cite{blondel2008fast}, suffer from bias of ignoring lowly-connected users. 
This is a type in which users with non-genuine retweets get assigned to significant communities for significant analysis, while users with low retweet rates get assigned to non-significant groups which leads to their exclusion from the study. On the other hand, some other community detection methods~\cite{6729613} suffer from low recall scores due to their exclusion of users from their belonging communities. Community detection methods with low recall scores can suffer from incorporating bias by excluding so many significant users. Although there are many different methods for the community detection task~\cite{Papadopoulos2012,gargi2011large}, we divided our analysis in two different sections: 
\begin{quote}
\begin{enumerate}
\item Methods that do not use node attributes for the community detection task.
\item Methods that use node attributes for the community detection task.
\end{enumerate}
\end{quote}
Despite the fact that there are different methods that can go under each of these two major groups discussed above, we picked the two most well known and broadly used approaches to conduct our analysis later in this paper.

\subsection{Non-Attributed Methods}
One of the widely used community detection methods is the Louvain method, which utilizes the modularity value to obtain partitions by optimizing an objective function discussed in~\cite{blondel2008fast}. However, this method tends to create small communities by putting introverted and low-degree users into small communities which can be biased towards low-degree people who are relevant by context to the significant groups but not connected through the network structure. The results of such instances will be discussed in detail in the results section of our proposed method, but we will give the reader a brief overview of such cases in this section. 
Placement of low-degree people into insignificant groups of their own may create the illusion that these small communities are communities that are not so relevant to the major groups that need to be studied and therefore can be ignored, but we observed that these communities had so much relevant and informative content to the study, based on our ground truth labels, that it is crucial to keep them in the significant communities. Merely low-degree rates and network structure should not cause a user to be ignored.
Low-connection in social networks may be as a result of various factors. Consider the retweet network as an example, low-connection may have different reasons, such as users having unique ways of expressing ideas yet relevant to the majority group's opinion, the tweet content being long which tends to repel other users to read and retweet them or in case of introverted users with small friendship circles tweets may not get enough attention or retweets. Network sampling can also change network measures and create unrealistic introverted users~\cite{gonzalez2014assessing,morstatter2013sample}. These reasons are not indicative of irrelevancy of topic and should not cause bias towards low-degree users of such types and their exclusion from significant communities.
%Low retweet rates can put these users into small introverted communities due to various reasons, such as the tweet content being long which tends to repel other users to read and retweet them or in case of introverted users with small friendship circles their tweets may not get enough attention or retweet rates, but this does not indicate that the user should not belong to one of the major communities. 
Instead what the modularity value does is to separate these users into introverted and small communities depriving them from being into their correct significant communities by only focusing on the network structure and its connections. Other methods of such type that do not utilize node attributes exist, such as BigCLAM and DEMON~\cite{yang2013overlapping,coscia2012demon}. Using node attributes in addition to the network structure will add additional information, as discussed in \cite{cho2016latent}, and can be helpful to reduce this type of biases in social networks. There are methods that use node attributes in combination with the network structure to create communities. In the next section, we will discuss some methods that utilize node attributes in addition to the network structure and discuss issues associated with them and the potential biases they create.

\subsection{Attributed Methods}
There are some attributed community detection methods discussed in \cite{Falih:2018:CDA:3184558.3191570}. From which, one of the widely known methods that uses node attributes in addition to the network structure is a generative model called CESNA~\cite{6729613}. This method restructures the network by incorporating node attributes. Experiments show that this method performs well by having a high precision value; however, it suffers from having a significantly low recall value. Low recall value may be indicative that this method also discards some of the important users which we are trying to avoid in order to minimize bias. In addition to CESNA, other methods have been proposed, such as Block-LDA~\cite{balasubramanyan2011block}, PAICAN~\cite{Bojchevski2018BayesianRA}, shared latent space models such as CLSM~\cite{cho2016latent}, and embedding based approaches like  LANE~\cite{huang2017label}, which utilize node attributes in order to detect communities, or ELAINE~\cite{Goyal:2018:ENE:3209542.3209571} which utilizes edge attributes. However, none of these methods tried to target the existing bias in social networks.

%% file: sections/datasets.tex
\vspace{-1pt}
\section{Obtaining Populations for Community Detection} 
The populations used in this study are drawn from social media with a particular focus on datasets with node attributes, such as text, and a social network structure. Moreover, we want datasets where the underlying communities come from different backgrounds. To satisfy this, we utilized two different datasets: Gamergate and U.S. Presidential Election. Both datasets have ground truth community labels. Our process for obtaining these labels will be discussed in detail.
\subsection{Gamergate}
The Gamergate dataset consists of tweets posted in 2014 between months of August through October. The tweets surround the Gamergate controversy~\cite{mortensen2018anger}. It contains 21,441 users who collectively produced 104,914 tweets. These users fall into one of the two groups surrounding the controversy. One group consists of Gamergate supporters who are tweeting about ethics in journalism and believe that regardless of the relationship between journalists and game developers, journalists should give honest reviews to game developers. %FM: Who is "them"? Could you clarify?
%NM: Fixed it!
The other group, Gamergate opposers, argues that Gamergate supporters attack female game developers and also feminist critics, and that they are not concerned with ethics in journalism, but are using the opportunity to attack women in the gaming industry.

\begin{figure}[H]
  \centering
  \includegraphics[width=0.46\textwidth]{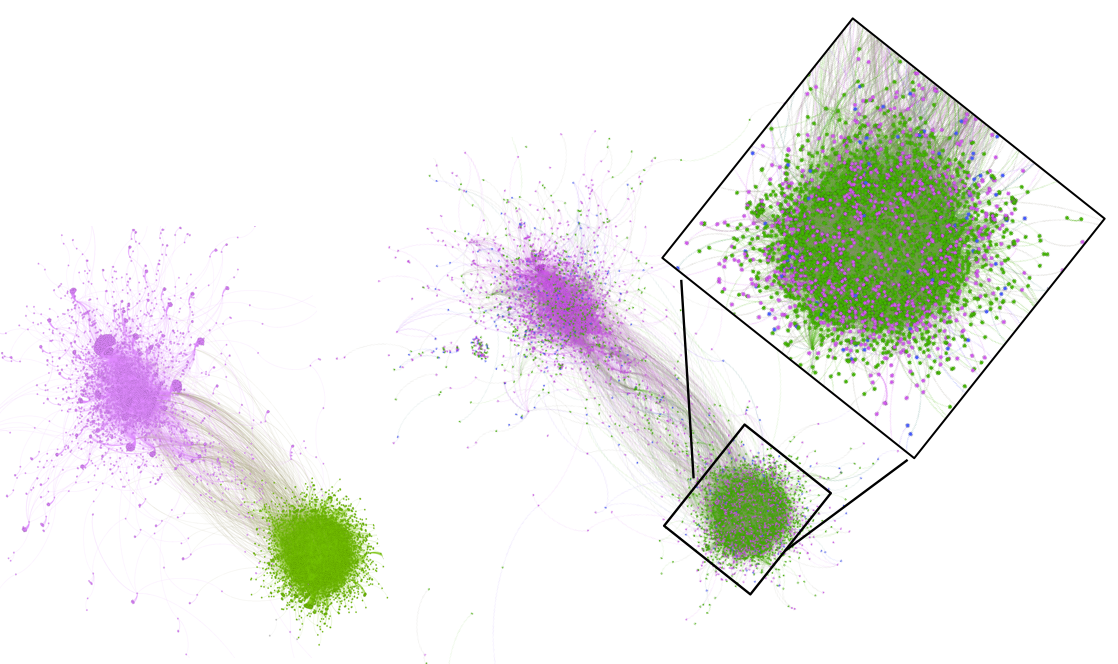}
  \caption{The Gamergate retweet network colored based on the network structure is shown on the left hand side, and the network colored by the ground truth labels is shown on the right hand side. The callout zooms one of the components, showing the disagreement between the two labeling approaches. Purple nodes represent Gamergate opposers and green nodes represent Gamergate supporters.}
  \label{Gamergatenetwork}
\end{figure}

In this study, we conducted an Amazon Mechanical Turk experiment discussed later in the paper to obtain ground truth labels for each of the users in this controversy. The retweet network of this dataset is shown in Figure \ref{Gamergatenetwork}.
\begin{figure}[H]
  \centering
  \includegraphics[width=0.5\textwidth,height=8cm]{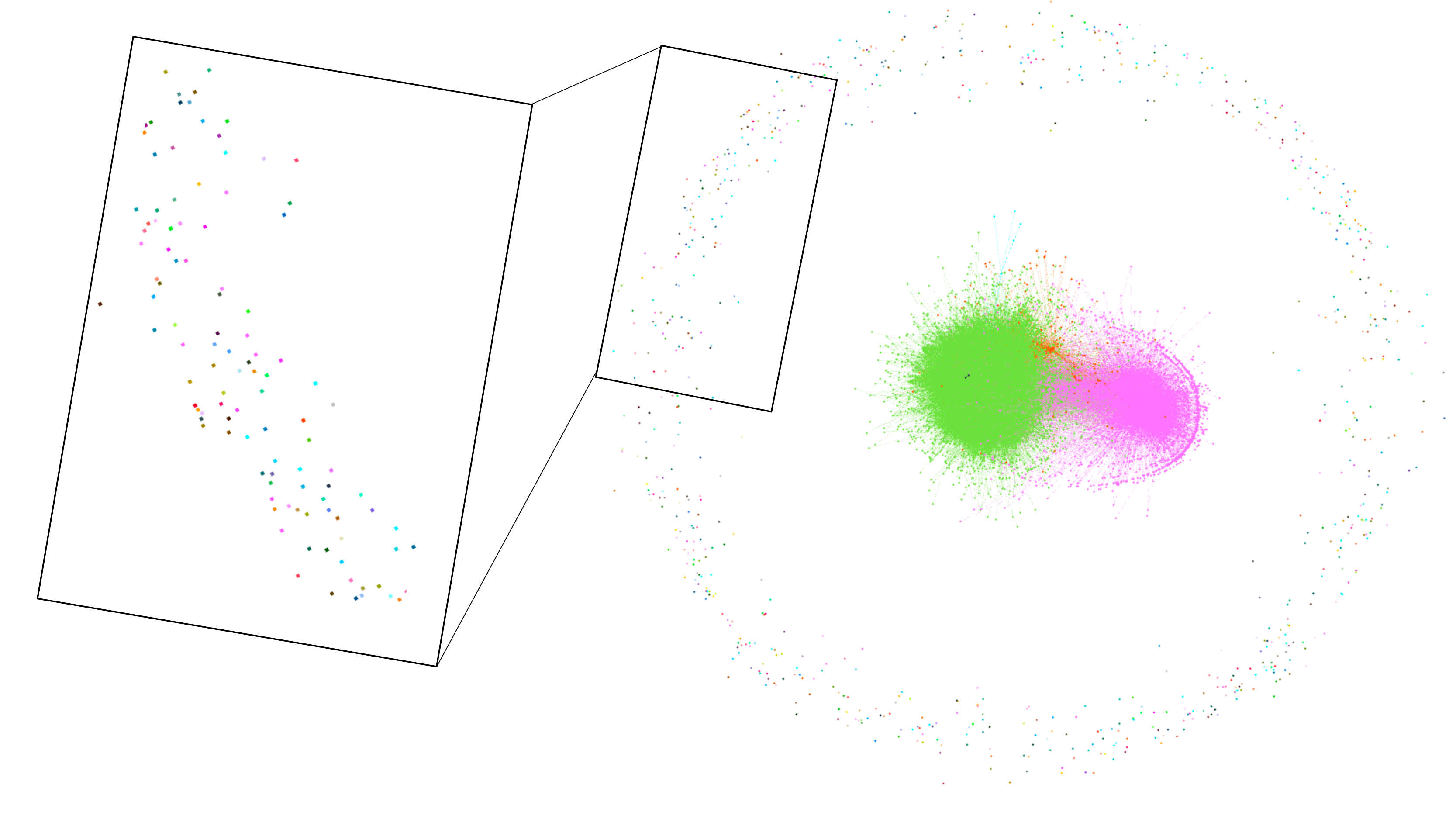}
  \caption{The 2016 U.S. presidential election seed-user retweet network colored by the  political party from modularity. The callout emphasizes the low-degree users.}
  \label{electionpic}
\end{figure}
\subsection{U.S. Presidential Election}
This dataset contains 10,074 users who discuss the U.S. presidential election in 2016. This dataset consists of two major groups which indicate the political party of each user. This dataset comes from~\cite{badawy2018characterizing} in which we only utilized the seed users from the whole dataset which brought our dataset size down from more than million users to 10,074 users since we required pure ground truth labels that were obtained away from the network structure and label propagation. The network structure of this dataset is shown in Figure \ref{electionpic}.  
\section{Obtaining Ground Truth Labels for Gamergate}
Unlike the 2016 presidential election dataset, the Gamergate dataset inherently lacks ground truth labels for each user. To obtain them, we conducted an Amazon Mechanical Turk experiment in which we asked human annotators to assign labels to each user in the Gamergate dataset.
\subsection{Experimental Setup}
In order to collect labels for each user in the Gamergate dataset, all the tweets associated to a user were mapped to the particular user so that the dataset was on a user level. Out of 21,441 total users, we excluded users who had only single tweets and duplicated users with same tweets. We then asked the turkers on Amazon Mechanical Turk to label each of the 8,128 users left based on their tweets into one of the following groups:
\begin{enumerate}
\item Gamergate Supporter: Users fighting for ethics in journalism and criticizing some women game-developers and journalists for their unethical relationships.
\item Gamergate Opposer:  Users advocating for women's rights and protecting attacks from Gamergate supporters to female game-developers and feminist advocates.
\item Unaffiliated: Users who belong to neither of the groups.
\end{enumerate}
Turkers were given complete description of the controversy and a detailed explanation of the labeling procedure. In order to make sure that the turkers were following the standards, some sanity check questions were put under each page for us to be able to identify bot turkers. These sanity check questions were trivial, made-up users with tweets that were easy to be categorized into one of the three groups, Gamergate Opposer, Gamergate Supporter, and Unaffiliated. After identifying bot turkers and excluding their labels from the total labels, we took the maximum agreement between 8,128 users that were labeled by at least three turkers. This resulted in a new dataset discussed in the next section which served as our ground truth.
\subsection{Dataset}
The Gamergate dataset had 21,441 users initially. In this study, we consider only users who posted at least two original tweets. The resulting dataset contains %After doing some preprocessing on this dataset by deleting the singleton users who had only one tweet % FM: How do we define singletons? We should be very careful here..  
%NM: fixe it!
8,128 users which were then labeled by the turkers. After analyzing the Turker agreement between 8,128 labeled users, we considered only users where turkers agreed two out of three times on a single label. After this filtering, we have 7,320 labeled users in total. These 7,320 users served as our ground truth labels.
\subsection{Results}
After obtaining the ground truth labels and having three ground truth groups of users, Gamergate supporters, Gamergate opposers, and unaffiliated, the communities obtained using network attributes were then compared with the groups obtained by the labels from the Mechanical Turk experiment. Surprisingly these results had a very low agreement which will be discussed in detail in "Community Detection Results" section. Figure \ref{electionpic} confirms this fact by showing the disagreement between the left hand side picture, which is colored based on the network structure, and the picture on the right colored by the ground truth labels. There is a significant amount of disagreement between these two results. The purple nodes represent Gamergate opposers and the green nodes represent Gamergate supporters. Using network structure and attributes would put almost all of Gamergate supporters in the green portion of the network and Gamergate opposers in the purple section of the network completely separated; however, the ground truth labels tend to have mixed users into each of the sections. The ground truth results are expected as many opposers may retweet Gamergate supporters; therefore, using network attributes merely on the retweet network might not be a good idea for separating these users, and other attributes and characteristics of users can be used for a more accurate community detection task. These results illustrate the fact that network does not explain everything and additional information is required in order to obtain accurate communities of users. This motivated us to come up with a new method that would not only address the bias from creating small communities and exclusion of low-degree nodes but would also have higher agreement with the ground truth communities by using other node attributes in addition to the network structure which will be discussed in the next section.

%% file: sections/approach.tex
\vspace{-1pt}
\section{Proposed Method}
% After observing flaws in the discussed methods through our experiments, we decided to come up with a method that would address the following flaws:
We have confirmed that ignoring low-degree users introduces bias into the resulting analysis of the data. In light of this, we propose a community detection approach that addresses the following issues:
\begin{algorithm}[H]
\caption{CLAN}
\begin{algorithmic} 
\STATE \textbf{Input:}
\STATE 1: Network
\STATE 2: Threshold
\STATE \textbf{Output:} Communities
\STATE  // Step 1: Finding the communities
\STATE $C$ = find\_communities(Network)
\STATE // Step 2: Classifying minority users into significant communities
\FOR{$C_i$ in $C$}
\IF{count($C_i$) $>$ Threshold}
\STATE Add $C_i$ to training set
\ELSE
\STATE Add $C_i$ to test set
\ENDIF

\STATE model = train(training set)
\ENDFOR
\STATE predictions = model.inference(test set)
\FOR{p in predictions}
\IF{p.label == $C_i$}
\STATE Add p.data to $C_i$
\ENDIF
\ENDFOR
\STATE return C
\end{algorithmic}
\end{algorithm}
\begin{quote}
\begin{enumerate}
\item Bias against low-degree nodes: Creating numerous small communities from putting introverted users in them results in the exclusion of those users from many data analysis tasks which creates a bias against these lowly-connected users. This bias is in favor of bots that would get potential falsified retweets and attention in the network, while it would exclude introverted users with less friends entirely.
\item Bias from low recall: Having low recall may mean exclusion of some users which is not desirable since we want to be able to maintain all the users for a more accurate study of the dataset no matter how hard it is to put them into their appropriate communities. We should not be able to only identify the popular users correctly but all the users regardless of their popularity in the network.
\item Higher predictive accuracy of the resulting communities. While addressing these issues, we simultaneously want to improve the accuracy of the communities discovered by our method. %Higher predictive accuracy of the 
%NM: Fixed this
\end{enumerate}
\end{quote}
We introduce CLAN: Communities from Lowly-connected Attributed Nodes which addresses the mentioned above issues as follows:
\begin{quote}
\begin{enumerate}
\item CLAN uses additional node attributes other than network attributes, such as text, to classify the lowly-connected users that were put into the small communities, those are the communities that are small in terms of containing users but large in number since they contain introverted and disconnected users who are mostly ignored due to the false belief that introverted users have irrelevant content in their text compared to the highly-connected users, into the significant communities that contain highly-connected users who get most of the attention in methods that solely focus on the network structure. %FM: You should define small and large.
%NM: fixed but I think the sentence is too wordy!!!!!!!!!!????????
\item By not changing the network structure and only adding information to this structure through the additional node attributes, such as text utilization, results obtained from CLAN would tend to have higher recall values.
\item From the reasons given in 1 and 2, we will show in our results through experimentation that CLAN has a superior performance and in a higher agreement with the ground truth communities.
\end{enumerate}
\end{quote}
CLAN uses node attributes to incorporate the introverted and lowly-connected users into their correct communities. Therefore, instead of creating small communities and putting these users into these insignificant communities by mistake, CLAN tends to utilize node attributes, in our case text attributes, to correctly put users into major communities that are significant to different down stream tasks. This would reduce the bias that will cause these users to be excluded from the data analysis tasks. In addition to that, by correctly utilizing these additional node attributes without changing the network structure, CLAN would obtain higher recall values and generally more accurate results that would be in a higher agreement with the ground truth communities while staying computationally more efficient.
The CLAN algorithm is a two step process, in which we first use unsupervised learning to develop communities using network attributes and modularity value. Once we have the communities, we would then turn the problem into a supervised classification problem in which we would classify the introverted users from insignificant communities into the major communities using additional node attributes that were held out in the first step, such as text attributes associated to a user. The first step of this process is a straightforward process in which we utilized the Louvain method discussed in~\cite{blondel2008fast}. Note that any community detection task that only utilizes the network, such as BigCLAM and DEMON~\cite{yang2013overlapping,coscia2012demon}, can replace this step which makes our algorithm flexible,general, and also capable of handling overlapping communities. 
Once this method is applied and the communities are obtained, we train a classifier on the majority communities using node attributes as features. The introvert users will then be classified into the majority groups that were used in our training process. The features can be any held out node features, such as text or hashtags that each user used in their tweets.
Unlike most of the generative methods that create new networks by combining network attributes and node attributes, we do not recreate the whole network structure, % FM: Let's not compare to CESNA so directly. What is meant by "recreate the whole network"? 
%NM: fixed this! Let me know if it is still unclear!
but we only add information to the existing network without changing its fundamental structure. This can also make our algorithm computationally more efficient than generative models which recreate the network.

%% file: sections/cd-expt.tex
\vspace{-1pt}
\section{Community Detection Results} \label{cd}
\begin{table*}
\begin{tabular}{ |p{3cm}||p{3cm}|p{3cm}||p{3cm}|p{3cm}|  }
 \hline
 \multicolumn{1}{|c||}{}&\multicolumn{2}{c||}{Gamergate Dataset}&\multicolumn{2}{c|}{U.S. Presidential Election Dataset}\\
 \hline
Method & F1 Score &Jaccard&F1 Score &Jaccard\\
 \hline
 CESNA   & 0.343    & 0.211&0.253 &0.149\\
 Modularity&  0.434   & 0.282 &0.753 &  0.604\\
 CLAN & \textbf{0.478}&\textbf{0.318} & \textbf{0.787}
&\textbf{0.649}\\
 \hline
\end{tabular}
\caption{The quantitative results obtained from calculating the F1 and Jaccard similarity scores with regards to the ground truth labels for each of the methods.}
\label{resulttable}
\end{table*}
\begin{table*}
\centering
\begin{tabular}{ |m{4.5cm}||m{4.5cm}||m{4.5cm}|  }
\hline \multicolumn{1}{|c}{}&\multicolumn{1}{c}{Unlabeled Users}&\multicolumn{1}{c|}{}\\
 \hline
 \multicolumn{1}{|c||}{Method}&\multicolumn{1}{c||}{Gamergate Dataset}&\multicolumn{1}{c|}{U.S. Presidential Election Dataset}\\
 %FM:  No need to say "percentage of" when you have the percents.
% FM: (Just minimized the table a little) Method & Unlabeled Users&Unlabeled Users\\
%NM:Fixed!
 \hline
 CESNA   &   69\%  &95\% \\
 Modularity& 21\%   &  20\%\\
 CLAN & 0\%&0\%\\
 \hline
\end{tabular}
\caption{Percentage of unlabeled users in each of the methods.}
\label{percentage} % (Add the label here)
\end{table*}
\begin{figure*}[!b]
  \centering
  \includegraphics[width=\textwidth]{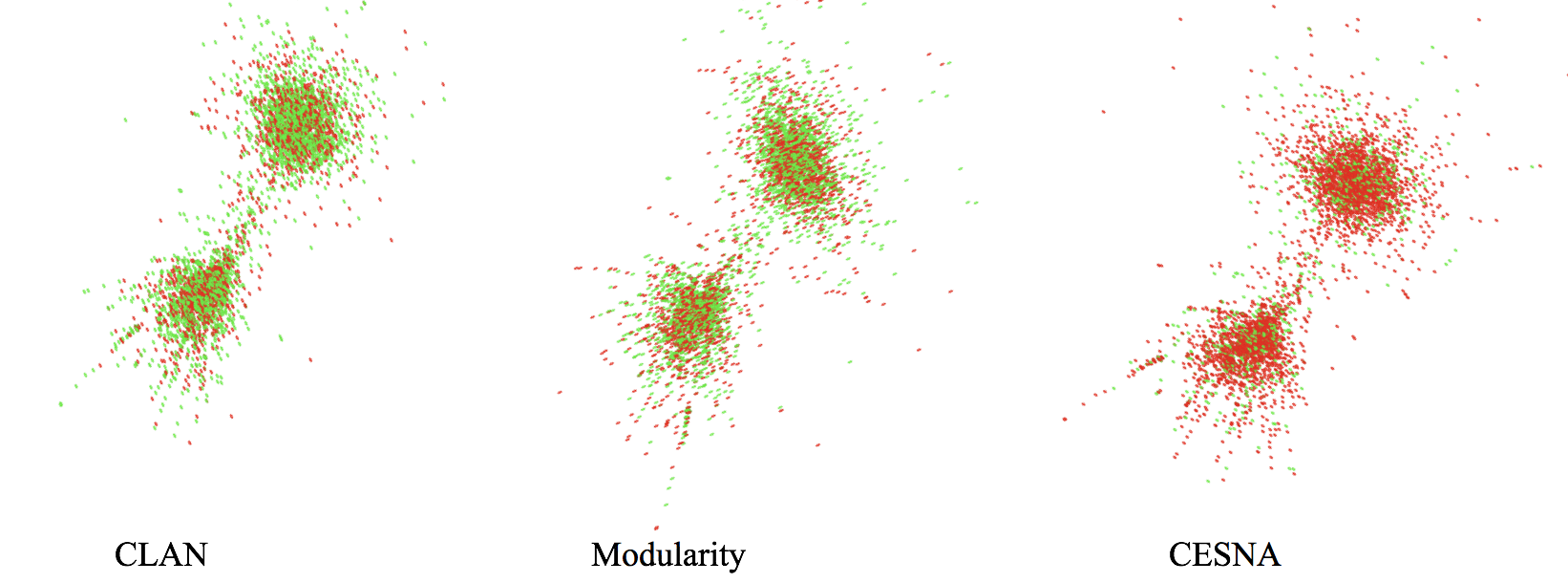}
  \caption{Networks colored by agreement with the ground truth labels for three methods for the Gamergate dataset.}
  \label{agreement}
\end{figure*}
 
\begin{table*}
\centering
\begin{tabular}{ |p{6.5cm}||p{2.5cm}|p{2.5cm}|p{2.5cm}|p{2.5cm}|  }
 \hline
 Gamergate Tweet&CESNA Label&Modularity Label&CLAN Label&Ground Truth Label\\
 \hline
``\#gamergate is like a particularly lame version of pakistani twitter political flame wars like ridiculously lame"&N/A&N/A&Anti Gamergate &Anti Gamergate\\[0.5pt]
 \hline
 ``\#gamergate is brutally put in its place by a journalist``&N/A&N/A&Anti Gamergate &Anti Gamergate\\[0.5pt]
 \hline
 ``\#gamergate \#done i suspect the \#ottawashooting has more in common with \#montrealmassacre \#columbine \#gamergate than it does with \#islam``&N/A&N/A&Anti Gamergate &Anti Gamergate\\[0.5pt]
  \hline
%  ``\#gamergate \#notyourshield a victory for \#gamergate or against yellow journalism \#notyourshield \#gamergate. gamers are like honey badgers you do not out-fight a honey badger  \#notyourshield \#gamergate may you receive a billion retweets. if \#notyourshield was recognized by the media ppl like joss whedon would have to admit they were wrong about \#gamergate  \#gamergate. read the calm demeanor of the gamergate person versus the fanaticism of the anti-gamergate ... where do you think gawker got their tactics from the \#gop and fox news practically taught them how to suppress \#gamergate``&N/A&N/A&Gamergate Supporter&Gamergate Supporter\\[0.5pt]
%   \hline
 ``\#gamergate i support \#gamergate and \#notyourshield i stand against harassment threats and doxxing no matter who or why``&N/A&N/A&Gamergate Supporter&Gamergate Supporter\\[0.5pt]
 \hline
 ``\#gamergate \#extralife2014 only two more days until our stream starts are you hyped we sure are current score to beat 100. I just supported \#gamergate extra life charity``&N/A&Gamergate Supporter&Gamergate Supporter&Gamergate Supporter\\[0.5pt]
 \hline
\end{tabular}
\caption{Sample Gamergate users with their corresponding tweets and labels assigned to each user by the three methods. ``N/A'' means that the method failed to classify the user.}
\label{gamergateexamples}
\end{table*}
\begin{table*}[!b]
\centering
\begin{tabular}{ |p{6.5cm}||p{2.5cm}|p{2.5cm}|p{2.5cm}|p{2.5cm}|  }
 \hline
 2016 U.S. Presidential Election Tweet & CESNA Label & Modularity Label & CLAN Label&Ground Truth Label\\
 \hline
``Trump wants to ban Muslim immigrants like my parents. I wrote a piece for telling him to go fuck himself.``&N/A &N/A &Democrat &Democrat\\[0.5pt]
 \hline
 ``Frauds at NY Times coverup Hillary's serious crimes...but try to make Trump into a criminal for LEGAL tax reduction strategy. WOW.``&N/A&Republican&Republican&Republican\\[0.5pt]
 \hline
 ``Fox News Host Makes Hillary Supporter Admit That Clinton Foundation Is A Huge Scam.``&Democrat&Republican&Republican&Republican\\[0.5pt]
  \hline
\end{tabular}
\caption{Sample 2016 U.S. Presidential Election users with their corresponding tweets and labels assigned by the three methods.}
\label{electionexample}
\end{table*}
Our goal is to report quantitative and qualitative results obtained through different experimentation in this paper. Hence, we use different visualizations and examples from datasets in hand, in addition to our numerical results to give the reader a better intuition of how our method, CLAN, performs compared to the existing state of the art. %FM: We need to rewrite this.
%NM: Rewrote this part and got rid of the superiority portion
\subsection{Evaluation Metrics}
For evaluation purposes, the F1 and Jaccard similarity scores are calculated for each of the datasets with respect to the ground truth labels. These scores are the average values of the total communities found in each of the datasets. 
%The ground truth labels and their collection procedure for each of the datasets are discussed in detail under the datasets section.%
% FM: Didn't we already do this?
%NM: True and I commented it out! Fixed.

\subsubsection{Quantitative Results} In this section, we would report quantitative and numerical results from our experiments. We will first report the scores for the F1 and Jaccard similarity scores between the ground truth labels and three different methods. We will show that CLAN outperforms other methods in terms of F1 and Jaccard similarity scores for both of the datasets. %FM: Isn't this redundant?
%NM: Fixed it!
Table \ref{resulttable} contains results for the F1 and Jaccard similarity scores obtained from comparisons done between the ground truth labels and labels obtained by applying each of the methods on the two datasets on hand.

In addition to reporting the F1 and Jaccard score results from the ground truth comparisons, we conducted another experiment in which we report the results indicating the number of the users that were labeled by each of the methods. % FM: we need to rewrite this.
%fixed
These numbers show the number of users that the method has excluded by not labeling them. This exclusion shows the bias of the method towards those users. Therefore, the more unlabeled users a method has the more susceptible to bias it is. In Table~\ref{percentage}, %FM: reference it like this. 
%NM:Fixed!
we reported the percentage of the users who were left unlabeled or put in insignificant communities in each of the methods from the two datasets. These results confirm the fact that our method, CLAN, has mitigated the bias towards the introverted users by incorporating them into the significant communities which would not be excluded from various down stream data analysis tasks.

The two sets of results reported in Tables \ref{resulttable} and \ref{percentage} confirm that not only is our method able to achieve superior predictive accuracy but also mitigate bias against introverted users by assigning them labels preventing them from exclusion.
%\FloatBarrier
\subsubsection{Qualitative Results} Besides the numerical results reported in the previous subsection, we want to further our analysis by showing some visualized results and real examples drawn from our datasets to further prove our results from the previous subsection. We will start our qualitative results by showing a visualization of the retweet network in the Gamergate dataset in the three methods discussed in this paper. Each node in these graphs represent a user and the nodes are color coded based on their agreement with the ground truth labels. The green nodes represent agreement between the label that was assigned to that particular user obtained from the method used and the ground truth label, and the red node represent disagreement between the two labels associated to that node. Therefore, more green nodes in a graph represent the degree of agreement of that method with the ground truth label and generally its superiority in terms of agreement with the ground truth compared to the other methods. The results of these visualizations are shown in Figure \ref{agreement}.
In addition to disagreement, the red nodes may also represent the fact that a method has low recall value and that many users were assigned no labels while the ground truth has assigned it a label. This of course is a sort of disagreement between the labels, so the nodes are colored as red. As expected, CESNA would have many red nodes as this method tends to have a very low recall value and as shown in Table \ref{percentage}, this method has a high tendency to exclude many users by not assigning them a label. Therefore, it suffers from low agreement with the ground truth and as expected highly covered by red nodes. This also confirms the existence of bias towards these red users who suffered from CESNA's low recall issue. The result associated to this method is shown on the far right side of Figure \ref{agreement}. As we move to the next method in the middle of Figure \ref{agreement}, we see less red nodes. This is because using modularity value has a higher recall and generally more agreement with the ground truth, but one can still spot many red users in this method. Moving on to the last graph on the far left side of Figure \ref{agreement}, we can see the graph associated to CLAN. Due to CLAN's use of node attributes in addition to the network attributes, it is able to obtain higher agreement with the ground truth and therefore less red nodes compared to the previous method. CLAN was able to address many red nodes in the top portion as well as the bottom portion of the graph compared to the modularity method.
\begin{figure*}
  \centering
  \includegraphics[width=\textwidth]{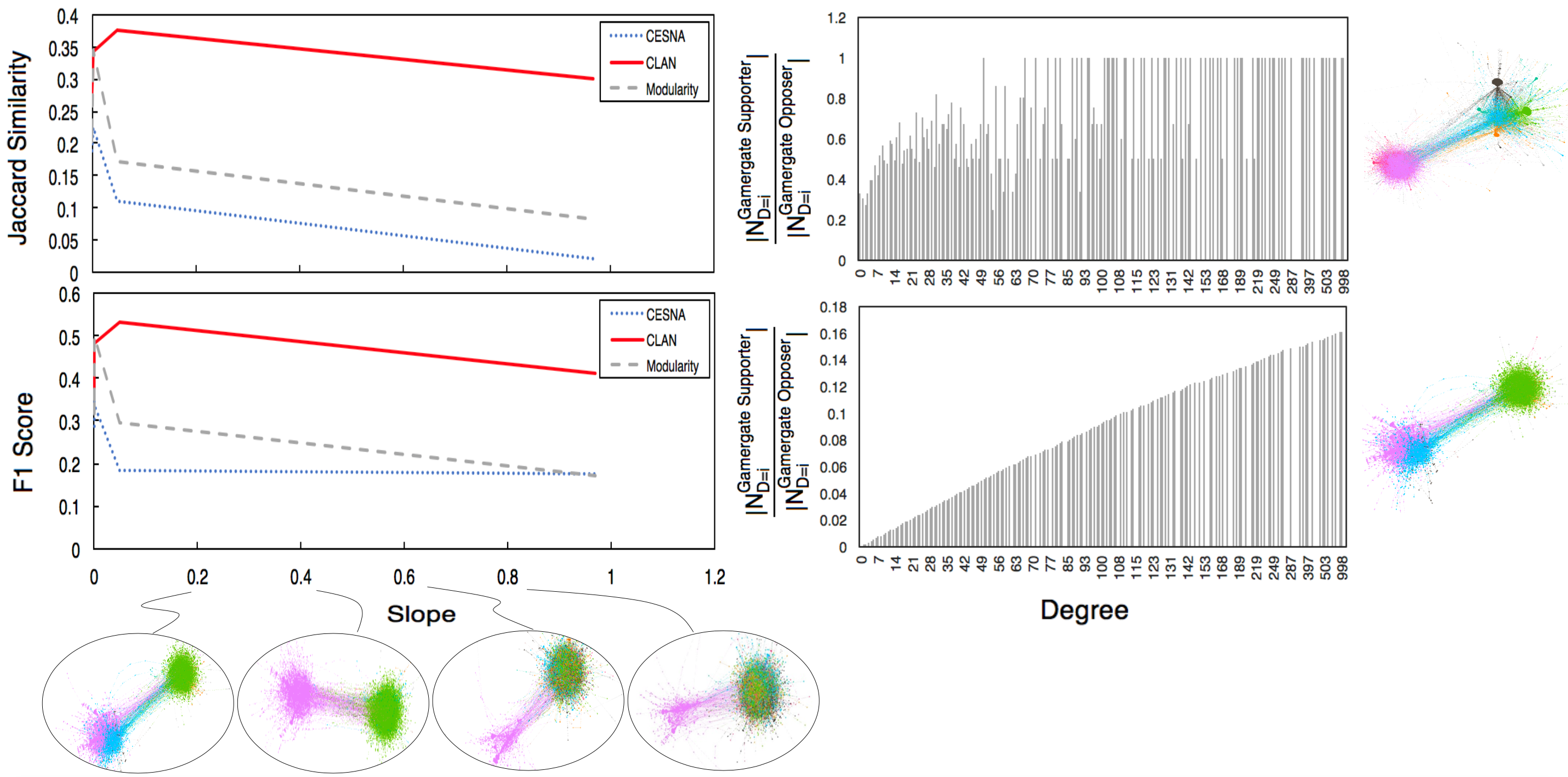}
  \caption{Synthetic distributions with their corresponding network and obtained results for the Gamergate dataset.}
  \label{synthetic1}
\end{figure*}
In addition to the agreement graphs shown in Figure \ref{agreement}, we provided two sets of tables, Table \ref{gamergateexamples} and \ref{electionexample}, in which some examples from each of the datasets are provided where it shows how each of the methods labeled a particular user with sets of tweets they tweeted. The ground truth labels are also listed for comparison purposes. Table \ref{gamergateexamples} contains the results from the Gamergate dataset, while Table \ref{electionexample} contains the results for the 2016 presidential election dataset. These examples also highlight the fact that the baseline approaches are suffering from the type of bias that roots from exclusion of users by not assigning them any labels or putting them into insignificant communities that would be excluded.

The qualitative results reported in this subsection also confirm the fact that the baseline methods have low agreement with the ground truth labels and suffer from bias towards low-degree and some users who are excluded from being labeled. The results from this and previous subsection also show the superiority of our method in terms of addressing these issues through various examples provided.

%% file: sections/synth-expt.tex
\section{CLAN's Resilience to Skewed Data}
% FM: We need to better motivate this experiment.
%NM: Changed it but not sure how effective it is.
One potential issue with community detection is that it could completely ignore entire communities \emph{if membership in that community is correlated with degree}. For example, a community of introverts is likely to be overlooked because they do not form many links. This presents a challenge to our approach as well, because Step 1 consists of identifying the major communities in the dataset. 
In order to test our method against this potential drawback, we conducted an experiment in which methods discussed in this paper would be tested in a synthetic environment in which the community distribution is skewed according to degree. This experiment serves two goals. First, it would compare the methods in terms of their performance on a different set of data. Secondly, it would show our method's behavior along with other methods while the distribution of each dataset is changed in terms of the node degree in the network which correlates with the popularity of a user in the social network.
\subsection{Experimental Setup}
In this experiment, we subsampled each dataset so that community membership is determined as a function of degree. Nodes with a particular degree are subsampled so that the statistic in Equation~\ref{equations} follows a particular trend. 
For example, in the Gamergate dataset, we changed the distribution such that the graph observed from plotting the degree vs. fraction of users in each of the two distinct groups, Gamergate supporter vs. Gamergate opposer, would have a particular trend. The same procedure was followed for the U.S. election dataset with its two groups being the political party a user was following. 
\begin{equation}
    \frac {\left|N_{D=i}^{Gamergate \, Supporter}\right|} {\left|N_{D=i}^{Gamergate \, Opposer}\right|}  \quad \textrm{and} \quad    \frac {\left|N_{D=i}^{Democrat}\right|} {\left|N_{D=i}^{Republican}\right|}
    \label{equations}
\end{equation} 

In Figure \ref{synthetic1}, the original distribution of the Gamergate dataset is shown on the top right corner of the figure with its corresponding network colored with the modularity value, and one of the synthetic distributions with a particular slope is shown in the bottom left with its corresponding network representation. Figure \ref{synthetic2} contains the same graphs and networks for the U.S. Presidential Election dataset. 

\subsection{Results}
The results for the Gamergate and U.S. Presidential Election datasets are  shown in Figures \ref{synthetic1} and \ref{synthetic2} respectively. The graphs located in the top left corner of the figures show the Jaccard similarity scores for each of the distributional settings, and the graph on the bottom left corner contains the results for the F1 scores.  The networks shown under each of the slope values in Figures \ref{synthetic1} and \ref{synthetic2} are the network of the users in the new distributional environments that have that particular slope range values. This confirms the fact that under different degree distributional settings CLAN can have reasonable and superior performance over the state of the art methods.

%% file: sections/conclusions.tex
\section{Conclusions and Future Work}
\begin{figure*}
  \centering
  \includegraphics[width=\textwidth]{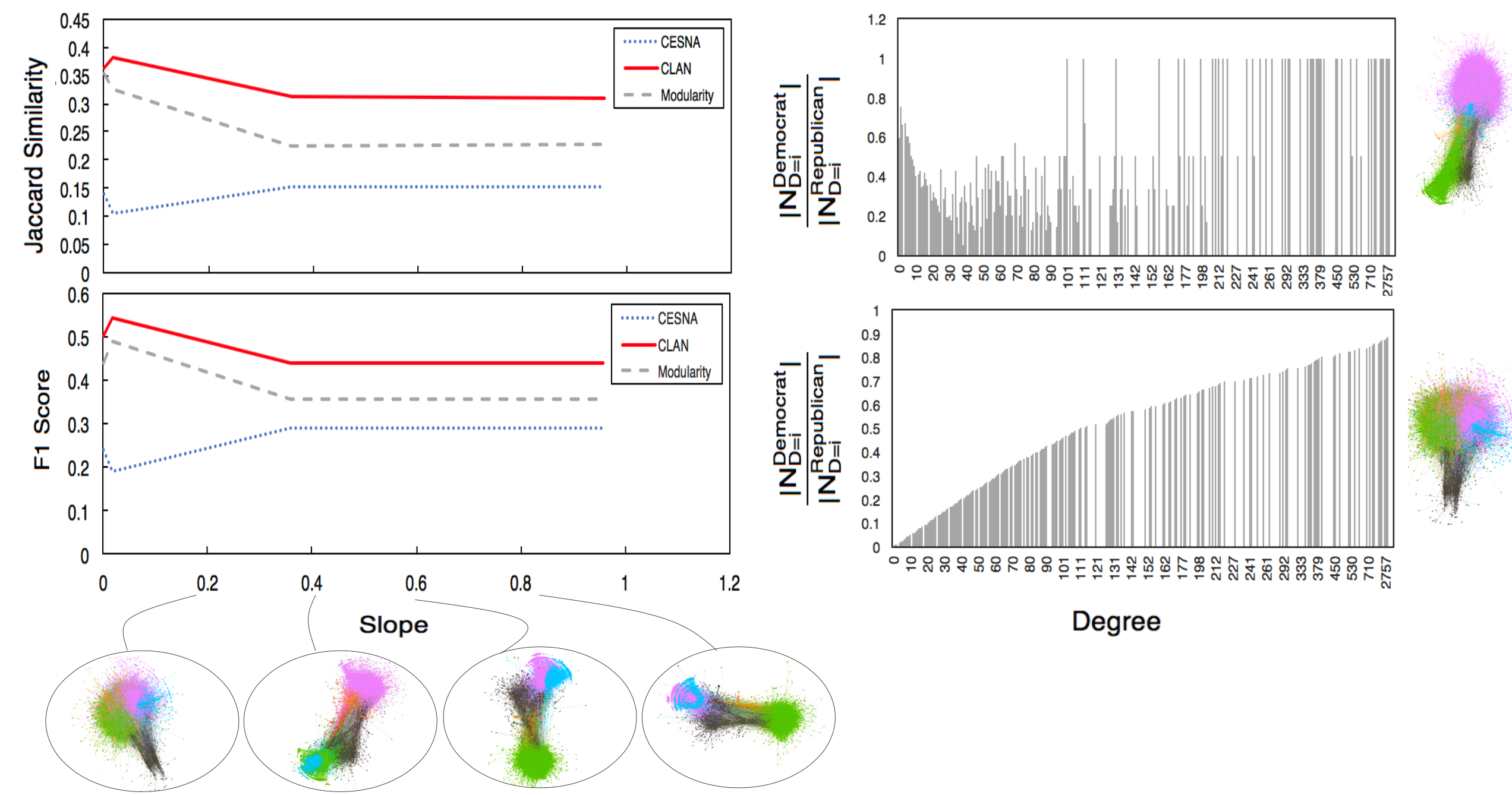}
  \caption{Synthetic distributions with their corresponding network and obtained results for the 2016 U.S. Presidential Election dataset.}
  \label{synthetic2}
\end{figure*}
In this paper, we introduce a new community detection method that mitigates the bias in existing detection methods that fail to properly account for sparsely-connected nodes in social networks. CLAN minimizes such biases by including the lowly-connected nodes into their true communities. Our empirical results demonstrate that inclusion of those users  enables CLAN to achieve overall superior performance in terms of F1-score and Jaccard similarity. We reported these results by providing evidence through our qualitative and quantitative experiments. Through qualitative analysis, we are able to show that these lowly-connected users, in aggregate, offer information that can be of use for analysis of social network data.

% Additionally, experiments were conducted in this paper both in terms of the ground truth label collection through human annotation and in terms of experiments done to prove CLAN's capability in outperforming the state of the art {\bf repetetive?}. We not only conducted experiments with the labels provided by the human annotators in our Gamergate dataset but also through automated methods in another dataset, the 2016 U.S. Presidential Election dataset, discussed in this paper. The results were then compared against these labels that were collected both through manual labor and automated approaches.

Finally, we show that our method is capable of outperforming other methods not only in real datasets but also in different types of synthetic environments with different population distributions, namely distributions where the users community is correlated with their connectivity. The results reported the performance of methods with regards to F1 and Jaccard similarity scores. 

We see two promising directions for future work. First, to extend this method to introduce a new hierarchical community detection method capable of detecting subgroups within larger communities. Second, we would like to consider edge attributes simultaneously with node attributes to enable this method to consider different types of connections.

\section{Acknowledgments}
This material is based upon work supported by the Defense Advanced Research Projects Agency (DARPA) under Agreement No. HR0011890019. We thank Greg Ver Steeg and Palash Goyal for helpful insights and discussions.

%% file: ms.bbl

\begin{thebibliography}{21}

%%% the Web address.

\ifx \showCODEN    \undefined \def \showCODEN     #1{\unskip}     \fi
\ifx \showDOI      \undefined \def \showDOI       #1{#1}\fi
\ifx \showISBNx    \undefined \def \showISBNx     #1{\unskip}     \fi
\ifx \showISBNxiii \undefined \def \showISBNxiii  #1{\unskip}     \fi
\ifx \showISSN     \undefined \def \showISSN      #1{\unskip}     \fi
\ifx \showLCCN     \undefined \def \showLCCN      #1{\unskip}     \fi
\ifx \shownote     \undefined \def \shownote      #1{#1}          \fi
\ifx \showarticletitle \undefined \def \showarticletitle #1{#1}   \fi
\ifx \showURL      \undefined \def \showURL       {\relax}        \fi
% invisible to TeX
\providecommand\bibfield[2]{#2}
\providecommand\bibinfo[2]{#2}
\providecommand\natexlab[1]{#1}
\providecommand\showeprint[2][]{arXiv:#2}

\bibitem[\protect\citeauthoryear{Badawy, Addawood, Lerman, and Ferrara}{Badawy
  et~al\mbox{.}}{2018}]%
        {badawy2018characterizing}
\bibfield{author}{\bibinfo{person}{Adam Badawy}, \bibinfo{person}{Aseel
  Addawood}, \bibinfo{person}{Kristina Lerman}, {and} \bibinfo{person}{Emilio
  Ferrara}.} \bibinfo{year}{2018}\natexlab{}.
\newblock \showarticletitle{Characterizing the 2016 Russian IRA Influence
  Campaign}.
\newblock \bibinfo{journal}{\emph{arXiv preprint arXiv:1812.01997}}
  (\bibinfo{year}{2018}).
\newblock


\bibitem[\protect\citeauthoryear{Balasubramanyan and Cohen}{Balasubramanyan and
  Cohen}{2011}]%
        {balasubramanyan2011block}
\bibfield{author}{\bibinfo{person}{Ramnath Balasubramanyan} {and}
  \bibinfo{person}{William~W Cohen}.} \bibinfo{year}{2011}\natexlab{}.
\newblock \showarticletitle{Block-LDA: Jointly modeling entity-annotated text
  and entity-entity links}. In \bibinfo{booktitle}{\emph{Proceedings of the
  2011 SIAM International Conference on Data Mining}}. SIAM,
  \bibinfo{pages}{450--461}.
\newblock


\bibitem[\protect\citeauthoryear{Blondel, Guillaume, Lambiotte, and
  Lefebvre}{Blondel et~al\mbox{.}}{2008}]%
        {blondel2008fast}
\bibfield{author}{\bibinfo{person}{Vincent~D Blondel},
  \bibinfo{person}{Jean-Loup Guillaume}, \bibinfo{person}{Renaud Lambiotte},
  {and} \bibinfo{person}{Etienne Lefebvre}.} \bibinfo{year}{2008}\natexlab{}.
\newblock \showarticletitle{Fast unfolding of communities in large networks}.
\newblock \bibinfo{journal}{\emph{Journal of statistical mechanics: theory and
  experiment}} \bibinfo{volume}{2008}, \bibinfo{number}{10}
  (\bibinfo{year}{2008}), \bibinfo{pages}{P10008}.
\newblock


\bibitem[\protect\citeauthoryear{Bojchevski and G{\"u}nnemann}{Bojchevski and
  G{\"u}nnemann}{2018}]%
        {Bojchevski2018BayesianRA}
\bibfield{author}{\bibinfo{person}{Aleksandar Bojchevski} {and}
  \bibinfo{person}{Stephan G{\"u}nnemann}.} \bibinfo{year}{2018}\natexlab{}.
\newblock \showarticletitle{Bayesian Robust Attributed Graph Clustering: Joint
  Learning of Partial Anomalies and Group Structure}. In
  \bibinfo{booktitle}{\emph{AAAI}}.
\newblock


\bibitem[\protect\citeauthoryear{Cho, Ver~Steeg, Ferrara, and Galstyan}{Cho
  et~al\mbox{.}}{2016}]%
        {cho2016latent}
\bibfield{author}{\bibinfo{person}{Yoon-Sik Cho}, \bibinfo{person}{Greg
  Ver~Steeg}, \bibinfo{person}{Emilio Ferrara}, {and} \bibinfo{person}{Aram
  Galstyan}.} \bibinfo{year}{2016}\natexlab{}.
\newblock \showarticletitle{Latent space model for multi-modal social data}. In
  \bibinfo{booktitle}{\emph{Proceedings of the 25th International Conference on
  World Wide Web}}. International World Wide Web Conferences Steering
  Committee, \bibinfo{pages}{447--458}.
\newblock


\bibitem[\protect\citeauthoryear{Coscia, Rossetti, Giannotti, and
  Pedreschi}{Coscia et~al\mbox{.}}{2012}]%
        {coscia2012demon}
\bibfield{author}{\bibinfo{person}{Michele Coscia}, \bibinfo{person}{Giulio
  Rossetti}, \bibinfo{person}{Fosca Giannotti}, {and} \bibinfo{person}{Dino
  Pedreschi}.} \bibinfo{year}{2012}\natexlab{}.
\newblock \showarticletitle{Demon: a local-first discovery method for
  overlapping communities}. In \bibinfo{booktitle}{\emph{Proceedings of the
  18th ACM SIGKDD international conference on Knowledge discovery and data
  mining}}. ACM, \bibinfo{pages}{615--623}.
\newblock


\bibitem[\protect\citeauthoryear{Diaz, Johnson, Lazar, Piper, and Gergle}{Diaz
  et~al\mbox{.}}{2018}]%
        {Diaz:2018:AAB:3173574.3173986}
\bibfield{author}{\bibinfo{person}{Mark Diaz}, \bibinfo{person}{Isaac Johnson},
  \bibinfo{person}{Amanda Lazar}, \bibinfo{person}{Anne~Marie Piper}, {and}
  \bibinfo{person}{Darren Gergle}.} \bibinfo{year}{2018}\natexlab{}.
\newblock \showarticletitle{Addressing Age-Related Bias in Sentiment Analysis}.
  In \bibinfo{booktitle}{\emph{Proceedings of the 2018 CHI Conference on Human
  Factors in Computing Systems}} \emph{(\bibinfo{series}{CHI '18})}.
  \bibinfo{publisher}{ACM}, \bibinfo{address}{New York, NY, USA}, Article
  \bibinfo{articleno}{412}, \bibinfo{numpages}{14}~pages.
\newblock
\showISBNx{978-1-4503-5620-6}
\urldef\tempurl%
\url{https://doi.org/10.1145/3173574.3173986}
\showDOI{\tempurl}


\bibitem[\protect\citeauthoryear{Falih, Grozavu, Kanawati, and Bennani}{Falih
  et~al\mbox{.}}{2018}]%
        {Falih:2018:CDA:3184558.3191570}
\bibfield{author}{\bibinfo{person}{Issam Falih}, \bibinfo{person}{Nistor
  Grozavu}, \bibinfo{person}{Rushed Kanawati}, {and}
  \bibinfo{person}{Youn\`{e}s Bennani}.} \bibinfo{year}{2018}\natexlab{}.
\newblock \showarticletitle{Community Detection in Attributed Network}. In
  \bibinfo{booktitle}{\emph{Companion Proceedings of the The Web Conference
  2018}} \emph{(\bibinfo{series}{WWW '18})}. \bibinfo{publisher}{International
  World Wide Web Conferences Steering Committee}, \bibinfo{address}{Republic
  and Canton of Geneva, Switzerland}, \bibinfo{pages}{1299--1306}.
\newblock
\showISBNx{978-1-4503-5640-4}
\urldef\tempurl%
\url{https://doi.org/10.1145/3184558.3191570}
\showDOI{\tempurl}


\bibitem[\protect\citeauthoryear{Gargi, Lu, Mirrokni, and Yoon}{Gargi
  et~al\mbox{.}}{2011}]%
        {gargi2011large}
\bibfield{author}{\bibinfo{person}{Ullas Gargi}, \bibinfo{person}{Wenjun Lu},
  \bibinfo{person}{Vahab Mirrokni}, {and} \bibinfo{person}{Sangho Yoon}.}
  \bibinfo{year}{2011}\natexlab{}.
\newblock \showarticletitle{Large-Scale Community Detection on YouTube for
  Topic Discovery and Exploration}. In \bibinfo{booktitle}{\emph{Fifth
  International AAAI Conference on Weblogs and Social Media}}.
\newblock


\bibitem[\protect\citeauthoryear{Gonz{\'a}lez-Bail{\'o}n, Wang, Rivero,
  Borge-Holthoefer, and Moreno}{Gonz{\'a}lez-Bail{\'o}n et~al\mbox{.}}{2014}]%
        {gonzalez2014assessing}
\bibfield{author}{\bibinfo{person}{Sandra Gonz{\'a}lez-Bail{\'o}n},
  \bibinfo{person}{Ning Wang}, \bibinfo{person}{Alejandro Rivero},
  \bibinfo{person}{Javier Borge-Holthoefer}, {and} \bibinfo{person}{Yamir
  Moreno}.} \bibinfo{year}{2014}\natexlab{}.
\newblock \showarticletitle{Assessing the bias in samples of large online
  networks}.
\newblock \bibinfo{journal}{\emph{Social Networks}}  \bibinfo{volume}{38}
  (\bibinfo{year}{2014}), \bibinfo{pages}{16--27}.
\newblock


\bibitem[\protect\citeauthoryear{Goyal, Hosseinmardi, Ferrara, and
  Galstyan}{Goyal et~al\mbox{.}}{2018}]%
        {Goyal:2018:ENE:3209542.3209571}
\bibfield{author}{\bibinfo{person}{Palash Goyal}, \bibinfo{person}{Homa
  Hosseinmardi}, \bibinfo{person}{Emilio Ferrara}, {and} \bibinfo{person}{Aram
  Galstyan}.} \bibinfo{year}{2018}\natexlab{}.
\newblock \showarticletitle{Embedding Networks with Edge Attributes}. In
  \bibinfo{booktitle}{\emph{Proceedings of the 29th on Hypertext and Social
  Media}} \emph{(\bibinfo{series}{HT '18})}. \bibinfo{publisher}{ACM},
  \bibinfo{address}{New York, NY, USA}, \bibinfo{pages}{38--42}.
\newblock
\showISBNx{978-1-4503-5427-1}
\urldef\tempurl%
\url{https://doi.org/10.1145/3209542.3209571}
\showDOI{\tempurl}


\bibitem[\protect\citeauthoryear{Huang, Li, and Hu}{Huang
  et~al\mbox{.}}{2017}]%
        {huang2017label}
\bibfield{author}{\bibinfo{person}{Xiao Huang}, \bibinfo{person}{Jundong Li},
  {and} \bibinfo{person}{Xia Hu}.} \bibinfo{year}{2017}\natexlab{}.
\newblock \showarticletitle{Label informed attributed network embedding}. In
  \bibinfo{booktitle}{\emph{Proceedings of the Tenth ACM International
  Conference on Web Search and Data Mining}}. ACM, \bibinfo{pages}{731--739}.
\newblock


\bibitem[\protect\citeauthoryear{Liao, Fu, and Strohmaier}{Liao
  et~al\mbox{.}}{2016}]%
        {Liao:2016:SUB:2858036.2858422}
\bibfield{author}{\bibinfo{person}{Q.~Vera Liao}, \bibinfo{person}{Wai-Tat Fu},
  {and} \bibinfo{person}{Markus Strohmaier}.} \bibinfo{year}{2016}\natexlab{}.
\newblock \showarticletitle{\#Snowden: Understanding Biases Introduced by
  Behavioral Differences of Opinion Groups on Social Media}. In
  \bibinfo{booktitle}{\emph{Proceedings of the 2016 CHI Conference on Human
  Factors in Computing Systems}} \emph{(\bibinfo{series}{CHI '16})}.
  \bibinfo{publisher}{ACM}, \bibinfo{address}{New York, NY, USA},
  \bibinfo{pages}{3352--3363}.
\newblock
\showISBNx{978-1-4503-3362-7}
\urldef\tempurl%
\url{https://doi.org/10.1145/2858036.2858422}
\showDOI{\tempurl}


\bibitem[\protect\citeauthoryear{Malliaros and Vazirgiannis}{Malliaros and
  Vazirgiannis}{2013}]%
        {malliaros2013clustering}
\bibfield{author}{\bibinfo{person}{Fragkiskos~D Malliaros} {and}
  \bibinfo{person}{Michalis Vazirgiannis}.} \bibinfo{year}{2013}\natexlab{}.
\newblock \showarticletitle{Clustering and community detection in directed
  networks: A survey}.
\newblock \bibinfo{journal}{\emph{Physics Reports}} \bibinfo{volume}{533},
  \bibinfo{number}{4} (\bibinfo{year}{2013}), \bibinfo{pages}{95--142}.
\newblock


\bibitem[\protect\citeauthoryear{Morstatter, Pfeffer, Liu, and
  Carley}{Morstatter et~al\mbox{.}}{2013}]%
        {morstatter2013sample}
\bibfield{author}{\bibinfo{person}{Fred Morstatter},
  \bibinfo{person}{J{\"u}rgen Pfeffer}, \bibinfo{person}{Huan Liu}, {and}
  \bibinfo{person}{Kathleen~M Carley}.} \bibinfo{year}{2013}\natexlab{}.
\newblock \showarticletitle{Is the sample good enough? Comparing data from
  twitter's streaming API with Twitter's firehose}. In
  \bibinfo{booktitle}{\emph{7th International AAAI Conference on Weblogs and
  Social Media, ICWSM 2013}}. AAAI press.
\newblock


\bibitem[\protect\citeauthoryear{Mortensen}{Mortensen}{2018}]%
        {mortensen2018anger}
\bibfield{author}{\bibinfo{person}{Torill~Elvira Mortensen}.}
  \bibinfo{year}{2018}\natexlab{}.
\newblock \showarticletitle{Anger, fear, and games: The long event of\#
  GamerGate}.
\newblock \bibinfo{journal}{\emph{Games and Culture}} \bibinfo{volume}{13},
  \bibinfo{number}{8} (\bibinfo{year}{2018}), \bibinfo{pages}{787--806}.
\newblock


\bibitem[\protect\citeauthoryear{Papadopoulos, Kompatsiaris, Vakali, and
  Spyridonos}{Papadopoulos et~al\mbox{.}}{2012}]%
        {Papadopoulos2012}
\bibfield{author}{\bibinfo{person}{Symeon Papadopoulos},
  \bibinfo{person}{Yiannis Kompatsiaris}, \bibinfo{person}{Athena Vakali},
  {and} \bibinfo{person}{Ploutarchos Spyridonos}.}
  \bibinfo{year}{2012}\natexlab{}.
\newblock \showarticletitle{Community detection in Social Media}.
\newblock \bibinfo{journal}{\emph{Data Mining and Knowledge Discovery}}
  \bibinfo{volume}{24}, \bibinfo{number}{3} (\bibinfo{date}{01 May}
  \bibinfo{year}{2012}), \bibinfo{pages}{515--554}.
\newblock
\showISSN{1573-756X}
\urldef\tempurl%
\url{https://doi.org/10.1007/s10618-011-0224-z}
\showDOI{\tempurl}


\bibitem[\protect\citeauthoryear{Ruan, Fuhry, and Parthasarathy}{Ruan
  et~al\mbox{.}}{2013}]%
        {ruan2013efficient}
\bibfield{author}{\bibinfo{person}{Yiye Ruan}, \bibinfo{person}{David Fuhry},
  {and} \bibinfo{person}{Srinivasan Parthasarathy}.}
  \bibinfo{year}{2013}\natexlab{}.
\newblock \showarticletitle{Efficient community detection in large networks
  using content and links}. In \bibinfo{booktitle}{\emph{Proceedings of the
  22nd international conference on World Wide Web}}. ACM,
  \bibinfo{pages}{1089--1098}.
\newblock


\bibitem[\protect\citeauthoryear{Tang and Liu}{Tang and Liu}{2010}]%
        {tang2010community}
\bibfield{author}{\bibinfo{person}{Lei Tang} {and} \bibinfo{person}{Huan Liu}.}
  \bibinfo{year}{2010}\natexlab{}.
\newblock \showarticletitle{Community detection and mining in social media}.
\newblock \bibinfo{journal}{\emph{Synthesis lectures on data mining and
  knowledge discovery}} \bibinfo{volume}{2}, \bibinfo{number}{1}
  (\bibinfo{year}{2010}), \bibinfo{pages}{1--137}.
\newblock


\bibitem[\protect\citeauthoryear{Yang and Leskovec}{Yang and Leskovec}{2013}]%
        {yang2013overlapping}
\bibfield{author}{\bibinfo{person}{Jaewon Yang} {and} \bibinfo{person}{Jure
  Leskovec}.} \bibinfo{year}{2013}\natexlab{}.
\newblock \showarticletitle{Overlapping community detection at scale: a
  nonnegative matrix factorization approach}. In
  \bibinfo{booktitle}{\emph{Proceedings of the sixth ACM international
  conference on Web search and data mining}}. ACM, \bibinfo{pages}{587--596}.
\newblock


\bibitem[\protect\citeauthoryear{Yang, McAuley, and Leskovec}{Yang
  et~al\mbox{.}}{2013}]%
        {6729613}
\bibfield{author}{\bibinfo{person}{J. Yang}, \bibinfo{person}{J. McAuley},
  {and} \bibinfo{person}{J. Leskovec}.} \bibinfo{year}{2013}\natexlab{}.
\newblock \showarticletitle{Community Detection in Networks with Node
  Attributes}. In \bibinfo{booktitle}{\emph{2013 IEEE 13th International
  Conference on Data Mining}}. \bibinfo{pages}{1151--1156}.
\newblock
\showISSN{1550-4786}
\urldef\tempurl%
\url{https://doi.org/10.1109/ICDM.2013.167}
\showDOI{\tempurl}


\end{thebibliography}
